\documentclass[10pt,hidelinks]{article}

\usepackage{algorithm}
\usepackage{amsthm}
\usepackage[noend]{algpseudocode}
\usepackage{array}
\usepackage{booktabs}
\usepackage{enumitem}
\usepackage{geometry}
\usepackage{graphicx}
\usepackage{hyperref}
\usepackage{listings}
\usepackage{mathtools}
\usepackage{multicol,multirow}
\usepackage[linewidth=1pt]{mdframed}
\usepackage{subcaption}
\usepackage{tabularx}
\usepackage{tcolorbox}
\usepackage{varwidth}

\newcommand{\myparagraph}[1]{\smallskip\noindent \textbf{\textit{#1.}}}

\graphicspath{{./images/}}

\newtheorem{theorem}{Theorem}[section]

\newtheorem{definition}{Definition}[section]

\newcommand{\cpp}{\texttt{C++}}

\algrenewcommand\algorithmicindent{1em}

\algnewcommand{\algcomment}[1]{\qquad{\color{blue}\emph{// #1}}}    % Trailing comment
\algnewcommand{\alglinecomment}[1]{{\color{blue}\emph{// #1}}}      % Full line comment

\algnewcommand\alglocal{\textbf{local }}                            % local variables
\algnewcommand\algreturn{\textbf{return }}                          % return statement
\algnewcommand\algeach{\textbf{each }}                              % 'each' for loops
\algnewcommand\algretire{\textbf{retire process}}                   % a job retires
\algnewcommand\algthis{\textbf{this}}                               % self reference

\algnewcommand\algorithmicpar{\textbf{par}}
\algnewcommand\algorithmicparfor{\textbf{parallel\_for}}
\algnewcommand\algorithmicwrite{\textbf{write}}
\algnewcommand\algmodassign{\textbf{write}}                         % Symbol for shared writes
\algnewcommand\algwritemod[2]{\algmodassign(#1, #2)}          		% Write to shared memory
\algnewcommand\algto{\textbf{to }}									% 'to' for loops
\algnewcommand\algis{\textbf{is}}                                   % 'is' for conditions
\algnewcommand\algnot{\textbf{not}}                                 % 'not' for conditions
\algnewcommand\algin{\textbf{in}}                                   % 'in' for containers
\algnewcommand\algempty{\textbf{empty}}                             % 'empty' for containers

\newcommand{\function}{\textbf{function}}

\algnewcommand\algand{\textbf{ and }}                               % 'and' for conditions
\algnewcommand\algor{\textbf{ or }}                                 % 'or' for conditions
\algnewcommand\algassign{\ensuremath{\gets} }                       % local variable assign

\algnewcommand\algtrue{\textbf{true}}                               % true literal
\algnewcommand\algfalse{\textbf{false}}                             % false literal
\algnewcommand\algnull{\ensuremath{\perp}}                          % null literal

\algnewcommand\algarray[1]{\textnormal{array}\ensuremath{\langle}#1\ensuremath{\rangle}}    % array type
\algnewcommand\algmod[1]{\textbf{mod}\ensuremath{\langle}#1\ensuremath{\rangle}}            % mod type

\algnewcommand\algorithmicwith{\textbf{with}}
\algnewcommand\algorithmicread{\textbf{read}}
\algnewcommand\algorithmicas{\textbf{as}}
\algblockdefx[READ]{Read}{EndRead}[2]
{\algorithmicwith\ \algorithmicread(#1)\ \algorithmicas\ #2\ \algorithmicdo}
{\algorithmicend\ \algorithmicread}

\makeatletter
\ifthenelse{\equal{\ALG@noend}{t}}%
{\algtext*{EndRead}}
{}%
\makeatother

\algnewcommand\algorithmicinparallel{\textbf{in parallel}}
\algblockdefx[PARFOR]{ParallelFor}{EndParallelFor}[1]
{\algorithmicfor\ #1\ \algorithmicdo\ \algorithmicinparallel}
{\algorithmicend\ \algorithmicfor}

\makeatletter
\ifthenelse{\equal{\ALG@noend}{t}}%
{\algtext*{EndParallelFor}}
{}%
\makeatother

\algnewcommand\algorithmicmod{\textbf{mod}}
\algnewcommand\algorithmicreadarray{\textbf{read\_array}}
\algnewcommand\algorithmicreadblock{\textbf{read\_block}}
\algnewcommand\algorithmicdynamicread{\textbf{dynamic\_read}}
\algnewcommand\algorithmicalloc{\textbf{alloc\_mod}}
\algnewcommand\algorithmicallocarray{\textbf{alloc\_array}}
\algnewcommand\algorithmicrun{\textbf{run}}
\algnewcommand\algorithmiccall{\textbf{call}}
\algnewcommand\algorithmicprop{\textbf{propagate}}
\algnewcommand\algorithmicexecute{\textbf{execute}}
\algnewcommand\doread{\textbf{do\_read}}

\newcommand{\algtab}{\hskip\algorithmicindent \hspace*{2em}}

\mathtoolsset{showonlyrefs}

\newenvironment{proofsketch}{%
  \renewcommand{\proofname}{Proof sketch}\proof}{\endproof}

\newcommand{\multipram}[0]{\emph{multiprefix} CRCW \ensuremath{\mathsf{PRAM}}\renewcommand{\multipram}[0]{multiprefix CRCW \ensuremath{\mathsf{PRAM}}}}

\newcolumntype{P}[1]{>{\centering\arraybackslash}p{#1}}
\newcolumntype{Y}{>{\centering\arraybackslash}X}

\begin{document}

  \title{Efficient Parallel Self-Adjusting Computation}
  
  \author{ \normalsize Daniel Anderson \\ \normalsize Carnegie Mellon University \\ \normalsize dlanders@cs.cmu.edu \and \normalsize Guy E. Blelloch \\ \normalsize Carnegie Mellon University \\ \normalsize guyb@cs.cmu.edu \and \normalsize Anubhav Baweja \\ \normalsize Carnegie Mellon University \\ \normalsize abaweja@cs.cmu.edu \and \normalsize Umut A. Acar \\ \normalsize Carnegie Mellon University \\ \normalsize umut@cs.cmu.edu}

%  \author{Daniel Anderson}
%  \affiliation{\institution{Carnegie Mellon University}}
%  \email{dlanders@cs.cmu.edu}
  
%  \author{Guy E. Blelloch}
%  \affiliation{\institution{Carnegie Mellon University}}
%  \email{guyb@cs.cmu.edu}
  
%  \author{Anubhav Baweja}
%  \affiliation{\institution{Carnegie Mellon University}}
%  \email{abaweja@@cs.cmu.edu}
  
%  \author{Umut A. Acar}
%  \affiliation{\institution{Carnegie Mellon University}}
%  \email{umut@cs.cmu.edu}
  
  \date{}

  \maketitle
  
  \begin{abstract}
  Self-adjusting computation is an approach for automatically producing dynamic algorithms from static ones.
The approach works by tracking control and data dependencies, and propagating changes through the dependencies
when making an update.
 Extensively studied in the sequential setting, some results on parallel self-adjusting computation exist, but are either only applicable to limited classes of computations, such as map-reduce, or are ad-hoc systems with no theoretical analysis of their performance. 

 In this paper, we present the first system for parallel self-adjusting computation that applies to a wide class of nested parallel algorithms and provides theoretical bounds on the work and span of the resulting dynamic algorithms.  As with bounds in the sequential setting, our bounds relate a ``distance'' measure between computations on different inputs to the cost of propagating an update.  However, here we also consider parallelism in the propagation cost.

The main innovation in the paper is in using Series-Parallel trees (SP trees) to track sequential and parallel control dependencies to allow propagation of changes to be applied safely in parallel.  We show both theoretically and through experiments that our system allows algorithms to produce updated results over large datasets significantly faster than from-scratch execution.
  We demonstrate our system with several example applications, including algorithms for dynamic sequences and dynamic trees. In all cases studied, we show that parallel
  self-adjusting computation can provide a significant benefit in both work savings
  and parallel time.
\end{abstract}

  \pagenumbering{gobble}  
  \clearpage
  \pagenumbering{arabic}
  
  \section{Introduction}

%\guy{Make sure we are consistent with self-adjusting vs
%  change-propagation.   Perhaps explain up-front, perhaps in a
%  footnote, that change propagation is an algorithm to implement
%  self-adjusting computation.}

Self-adjusting computation is an approach to automatically, or semi
automatically, convert a (suitable) static algorithm to a dynamic
one~\cite{DemersReTe81,pughte89,acar2002adaptive,acarblha03-abbrv,acar2004dynamizing,acar05,hammeracch09,chenacta14}. Most often, self-adjusting computation
is implemented in the form of a \emph{change propagation} algorithm.
The idea, roughly, is to run a static algorithm while keeping track
of data dependencies.  Then when an input changes (e.g. adding an edge
to a graph), the change can be propagated through the computation,
updating intermediate values, creating new dependencies, and updating
the final output.  Not all algorithms are suitable for the
approach---for some, updating a single input value could propagate
changes through most of the computation.  To account for how much
computation needs to be rerun, researchers have studied the notion of
``stability''~\cite{acar2004dynamizing,acar05} over classes of
changes.  The goal is to bound the ``distance'' between executions of
a program on different inputs based on the distance between the
inputs.  For example, for an appropriate sorting algorithm adding an
element to the unsorted input list ideally would cause at most
$O(\log n)$ recomputation, and that recomputation could be propagated
with a constant factor overhead.  This would lead to the performance
of a dynamic binary search tree.

In the sequential setting this approach has been applied to a wide
variety of algorithms, with various bounds on the stability, and also
cost of change propagation as a function of the computational distance.
Applications includes dynamic trees~\cite{acar2004dynamizing}, kinetic data
structures~\cite{acarbltavi06,acarbltatu08,acarhutu11}, convex hulls~\cite{acar2009experimental}, Huffman
coding~\cite{acarblletatu10}, and Bayesian inference~\cite{acarihmesu07}.

More recent work~\cite{acar2020batch,Bhatotia11,burckhardtlesayiba11,hammer2007proposal,bhatotia2015ithreads}
has studied applying change propagation in parallel,
allowing for batch dynamic updates---e.g., adding a set of edges to an
existing graph and then propagating those changes in parallel.  Batch
updates are particularly important in practice due to the rapid rate
of modifications to very large data sets such as the web graph or social
networks.  Furthermore, in principle, parallelism and change
propagation should work well together since algorithms with shallow
dependence chains tend both to be good for parallelism (since fewer
dependencies means more task can run in parallel) and for dynamic
updates (since changes will not have to propagate as deeply).  Indeed
several researchers have studied the approach and developed systems in
the applied setting, which show good performance
improvements~\cite{Bhatotia11,burckhardtlesayiba11,bhatotia2015ithreads}
on tasks such as map-reduce.

In the theoretical setting, recent work has studied bounds on
the cost of change propagation for a class of so-called
``round-synchronous'' computations~\cite{acar2020batch}.  This was
applied to generate efficient algorithms for batch-dynamic trees,
supporting batches of links and cuts among other operations.  However,
the round-synchronous nature limits the applicability to certain algorithms that fit the restrictive model.

In this paper we develop a more general framework for supporting
self-adjusting computation for arbitrary nested-parallel algorithms.   We
prove bounds on the cost of change propagation in the framework based
on an appropriately defined distance metric.    We have also
implemented the framework and run experiments on a variety of
benchmarks.
A nested parallel program is one that is built from
arbitrary sequential and parallel composition.  A computation is
defined recursively as either two computations that are composed in
parallel (a fork), two that are composed sequentially, or the base
case which is a sequential \emph{strand}.
Multiway forking can easily be
implemented by nesting parallel compositions.
%This model is supported
%by parallel languages and libraries such as Cilk~\cite{Cilk11}, Java
%Fork-join~\cite{Lea00}, OpenMP~\cite{Dagum98}, and Intel's TBB~\cite{reinders2007intel}.

The crux of our technique is to represent a computation by a
dependency graph that is anchored on a \emph{Series-Parallel tree}
\cite{feng1999efficient}, or \emph{SP tree} for short.
An SP tree corresponds to the sequential and parallel composition of
binary nested parallel programs---i.e., parallel composition consists
of a $P$ node with two children (the left-right order does not
matter), and sequential composition consists of an $S$ node with two
children (here the order does matter).  The leaves are sequential
strands of computation, and can just be modeled as leaf $S$
nodes.   The SP tree represents the control dependencies in the
program---i.e., that a particular strand needs to executed before
another strand.
We additionally introduce R nodes to indicate data reads,
which are used to track data dependencies between writes and
reads---i.e., that a particular read depends on the value of a
particular write.
Together we refer to the trees as \emph{RSP trees}, corresponding to
the three types of nodes.
The RSP tree of a computation allows propagating a change in a way
that respects sequential control dependencies while allowing
parallelism where there is no dependence.  We prove that a parallel change propagation algorithm can propagate changes through the computation in a manner that is both efficient and scalable.

Programs written in our framework write their inputs and any non-local
values that depend on them into ``modifiable references'', or
\emph{modifiables} for short, which track all reads to them and
facilitate change propagation. Like previous work on sequential change
propagation~\cite{acar2004dynamizing}, we achieve our 
efficiency by restricting input programs to those which write to each
modifiable exactly once. All race-free functional programs satisfy
this restriction. We note that since local variables do not need to be
tracked, they are not bound by this restriction, so the scope of
programs amenable to our framework is not just those which are
purely functional. 
% In this sense, our framework is actually less
% general than the previous work of Burckhardt et
% al.~\cite{burckhardtlesayiba11}, which supports multiple writes to the
% same variable. However, we show that the added simplicity of the
% write-once restriction allows our framework to obtain
% theoretically-efficient bounds and provable correctness guarantees
% without sacrificing the ability to express a range of useful
% algorithms.

Roughly speaking, given two executions of the same algorithm on
different inputs, we define the \emph{computation distance} to be the
work that is performed by one but not the other (see
Definition~\ref{def:distance} for the full definition). We then show
the following theorem that bounds the runtime of the change
propagation algorithm as a function of computation distance.

\begin{theorem}[Efficiency]\label{thm:change-prop-fast}
  Consider an algorithm $A$, two input states $I$ and $I'$, and their corresponding RSP trees $T$ and $T'$. Let $W_\Delta = \delta(T,T')$ denote the computation distance, $R_\Delta$ denote the number of affected reads, $s$ denote the span of $A$, and $h$ denote the maximum heights of $T$ and $T'$. Then, change propagation on $T$ with the dynamic update $(I, I')$ runs in $O(W_\Delta + R_\Delta \cdot h )$ work in expectation and $O(s\cdot h)$ span w.h.p.\footnote{We say that an algorithm has $O(f(n))$ cost \emph{with high probability (w.h.p.)}
    if it has $O(c\cdot f(n))$ cost with probability at least
    $1 - 1/n^{c}$, $c \geq 1$.  }
\end{theorem}

\noindent We have implemented the proposed techniques in a library for C++, which
we call PSAC++ (Parallel Self-Adjusting Computation in C++).
The library allows writing parallel self-adjusting programs by using
several small annotations in a style similar to writing 
conventional parallel programs.
Self-adjusting programs can
respond to changes to their data by updating their output via the
built-in change propagation.
Our experiments with several applications show that parallel change propagation can handle a broad
variety of batch changes to input data efficiently and in a scalable
fashion.
For small changes, parallel change propagation can
yield very significant savings in work; such
savings can amount to orders of magnitude of improvement.
For larger changes, parallel change propagation may save some work, and still
exploit parallelism, yielding improvements due to both reduction in
work and an increase in scalability. We summarize the contributions of the paper
as follows:
\begin{itemize}[leftmargin=10pt]
  \item  A general approach for parallel change propagation based on
using RSP trees to safely propagate changes in the
correct order while allowing parallelism in the propagation.

\item Theoretical bounds on the work (sequential time) and span
  (parallel time) of our algorithms.

\item An implementation as C++ library, with six example applications
  to study as benchmarks.

\item
  Experimental results that confirm what is backed up by our theoretical
  analysis, that parallel change propagation is efficient for a range
  of applications.
\end{itemize}

\subsection{Technical Overview}

The idea of the change propagation algorithm is first to run
an algorithm on some initial input while keeping a trace of reads and
writes to ``non local'' locations.  This trace can be thought of as a
write-read dependence graph, indicating what reads depend on what
writes (also called a data dependence graph).  Along with each read
the trace also stores the code that was run on the value, and
maintains some form of control ordering of the execution.  When an
input is updated at particular locations, the change propagation
algorithm knows what read those locations and reruns them.  This can
cause new reads and writes that both update the trace, and create
changes that have to propagated to their readers.  Importantly, and
one of the biggest challenges in change propagation, is that the reads
that rerun have to do so in control order, otherwise they
could use stale information.  For example, if a read A, and a later a
read B in program control order both need to be rerun, running B first
might miss updates by A.  Since A could do something different when
rerun, the trace might not even know there will be a data dependence
between them (A is now going to write to something B reads).  This
means that the topological order on the trace's data dependence graph
in insufficient for safety, and that control dependencies also
need to be considered.

In the sequential setting, the total order of all instructions is
typically maintained using a dynamic list-maintenance data
structure~\cite{dietzsl87} keeping all reads in time order.  The
structure needs to be dynamic since during propagation new computation
can be added, and old deleted, at arbitrary points in the
ordering.  During change-propagation, all reads that are affected by a
write are placed in a priority queue prioritized by this order, and always
processing the earliest first.  

For our work on parallel change propagation, the broad idea is to
organize the control dependencies of the program around the RSP tree.  Unlike
the sequential case, instead of having a total order of execution
time, the RSP tree effectively keeps track of the parallel partial
order of control dependencies among the strands.  As with the sequential
case, we also keep track of all write-read data dependencies. 
% It is useful to
% separate in one's mind dependencies from program order (i.e. the RSP
% tree), and the write-read dependencies, which cross between nodes of
% the RSP tree. 
Unlike the sequential algorithm which uses a priority
queue of time order, our algorithm instead uses the RSP tree itself to
maintain the partial order among strands---and this allows running multiple tasks
in parallel during the propagation.

The initial run builds the RSP tree.  It stores on each read (R) node
a closure to rerun if the value it read changes\footnote{A closure is
  a code pointer along with needed local variables.}.  When the input
is modified, the change-propagation algorithm identifies all readers
of the changed values.  We refer to these as the \emph{affected readers}. Now,
instead of adding them to a priority queue by sequential time order,
the algorithm makes some markings in the RSP tree.  In particular it
starts at each affected reader and marks all ancestors in the RSP
tree.  It then traverses the RSP tree and using the marks finds readers that require and
are safe to rerun, i.e., only descending if
a node is marked.  Whenever it gets to a $P$ node, the 
algorithm traverses down whichever children are marked (either left,
right, or both in parallel), and whenever it gets to an internal $S$
node, it traverses down the left branch if it is marked, and then the
right branch if it is marked.  At an $S$ node the algorithm never goes
down both branches simultaneously since that would be unsafe---it
would run strands in parallel that should be run sequentially.

Whenever the traversal meets an affected reader,
change propagation runs
the closure associated with the reader and updates the resulting
computation and its corresponding subtree of the RSP tree, possibly
cascading new reads and writes and marking additional regions of the RSP tree
for additional propagation. Once the marked regions of the tree have all been
traversed, change propagation is complete and the computation will be
fully up to date.

  %\section{Preliminaries}

\subsection{Related Work}

While there is little work on fully fledged parallel self-adjusting computation, a few systems have been developed.

\myparagraph{Incoop} On the programming language side, Bhatodia et al.~\cite{Bhatotia11} develop a framework
for self-adjusting computation in the map-reduce paradigm. Their system specifically targets map-reduce-style computations in a distributed model of computation, and does not provide any theoretical guarantees on the runtime of updates.

\myparagraph{Two for the price of one}
Burckhartd et al.\ \cite{burckhardtlesayiba11} were the first to develop a general-purpose system for parallel self-adjusting computation. They do so by extending the so-called concurrent
revisions model with primitives for self-adjusting computation.
This model enables programs to express fork-join parallelism with \emph{versioned}
types that allow multiple threads to concurrently write to an object
that is automatically aggregated at the join point, in a style similar
to Cilk reducers~\cite{frigo2009reducers}. Their algorithm for self-adjusting
computation then essentially performs memoisation of the versioned writes done by
each fork, allowing them to be looked up and re-used if their dependencies
haven't changed. Their work is evaluated on a set of five benchmark
problems, where it is demonstrated experimentally that the combination
of parallelism and self-adjusting computation is worthwhile, producing
both work savings and parallel time speedups. This work, however, is purely
experiential and does not provide theoretical guarantees on the
runtime of updates.
%Indeed, it appears that in the worst case, their read operation could take time proportional to the size of the computation.

\myparagraph{iThreads}
Bhatodia et al.~\cite{bhatotia2015ithreads} develop iThreads, a pthreads drop-in replacement that automatically dynamizes the underlying program. The advantage of such a system is that dynamization is completely automatic; the programmer does not even have to annotate their code or add additional primitives to perform self-adjusting computation. The corresponding downside is that the dynamization is very coarse grained, since the only units of work that can be re-executed are the entire pthread computations. The user is therefore unable to fine tune the dynamism, which is often important to optimize self-adjusting programs. Theoretical guarantees on the runtime of updates are not provided.

\myparagraph{PAL} Hammer et al.~\cite{hammer2007proposal} present Parallel Adaptive Language (PAL), a proposed (though not fully implemented) language for parallel self-adjusting computation. Like other works, including ours, it represents the trace of a computation using a tree structure, which encodes parallel and sequential dependencies between computations and the data that they read. Their proposed algorithm, however, requires several nontrivial data structures, some of which, such as an efficient concurrent fully dynamic lowest common ancestor (LCA) structure, do not yet exist. Their evaluation therefore consists of a simulation of the work that would be performed by the algorithm if said data structures were available, rather than a realistic evaluation.

\subsection{Model of Computation}

We analyze algorithms in the \emph{work-span} model,
  where work is the total number of instructions performed by the algorithm and span (also called depth) is the length of the longest chain of sequentially dependent instructions \cite{Blelloch96}. The model can work-efficiently cross simulate the classic CRCW PRAM model~\cite{Blelloch96}, and the more recent Binary Forking model~\cite{blelloch2020optimal}, incurring at most an additional $O(\log^*(n))$ factor overhead in the depth due to load balancing. An algorithm with work $W$ and span $s$ can be ran on a $p$-processor PRAM in $O(W/p + s)$ time~\cite{Brent74}.

  \section{Framework}

Our framework for parallel self-adjusting computation is built around a
set of core primitives that are easy to integrate into existing
algorithms. In this section, we describe these primitives
and give an example algorithm for illustration.

\begin{figure}[h]
  \centering
  \small
  \begin{mdframed}
    \algorithmicwrite{}(dest: $\alpha\ \algorithmicmod{}$, value : $\alpha$)\\
    \algorithmicalloc{}(T: \textbf{type}) : T \algorithmicmod{}\\
    \algorithmicread(m : $(\alpha_1\ \algorithmicmod, ..., \alpha_k\ \algorithmicmod)$, r : $\alpha_1 \times ... \times \alpha_k \mapsto ()$)\\
    \algorithmicpar{}(left\_f : $() \mapsto ()$ , right\_f : $() \mapsto ()$)\\
    \algorithmicrun{}($f : () \mapsto ()$) : $S$\\
    \algorithmicprop{}(root : $S$)
  \end{mdframed}
  \caption{Interface for Parallel Self-Adjusting Computation}
\end{figure}

\myparagraph{Modifiables}
The primary mechanism by which computations are dynamised
is through the use of \emph{modifiable} variables. A modifiable variable,
or modifiable for short, is either a value that is part of the input
to the algorithm, or a nonlocal variable whose value depends on the value of
another modifiable variable. Algorithms are dynamised
by placing their inputs in modifiables, and ensuring that all nonlocal variables
whose values depend on a modifiable are also placed in modifiables.
When a modifiable is updated, our framework
can then automatically determine which values are affected by the
resulting changes, and automatically propagate the appropriate updates.

Modifiables
can be allocated either statically, i.e.\ before the computation is run,
or dynamically, in which case their lifetime will be tied to the scope of
the computation that allocated them. Writing to modifiables is achieved
using the \algorithmicwrite{} operation. 
We require that each modifiable is written to at most once during each
run of the computation, and that modifiables are not read before they are written.
We also require that modifiables are only read from and written to by computations
that are in the dynamic (nested) scope of the computation that allocated it.

\myparagraph{Read operations} To ensure that dependencies are tracked, modifiables must be
read using the \algorithmicread{} operation. \algorithmicread{} reads the values of the given modifiables and invoke the given reader function with their current values as arguments.

\myparagraph{Parallelism} We support fork-join parallelism through
a binary fork operation \algorithmicpar{}, which takes two code thunks and executes them in parallel.

\myparagraph{Control} Computations are initiated with the \algorithmicrun{} operation, which returns a handle to the computation (represented by the root of the RSP tree). After making changes
to the input, changes are propagated using the \algorithmicprop{} operation.

\myparagraph{Additional primitives} For convenience, we also support an \algorithmicallocarray{} operation
and a corresponding \algorithmicreadarray{} operation for allocating and reading arrays of modifiables.
We also support a \mbox{\algorithmicparfor{}} primitive, which executes a given function over a range of values in
parallel. We omit the details of these primitives.

\myparagraph{A note on randomness} We require that all algorithms implemented in our framework be deterministic.
That is, given some input, if re-executed they must produce exactly the same output. It is still possible, and
indeed we have several in our application examples, to implement randomized algorithms. To do so, the randomness
must simply be pre-generated before executing the computation to ensure that, when re-executed, the same results
will be obtained.

\myparagraph{Example} To illustrate our framework, we give an implementation of a parallel divide-and-conquer sum function. See Algorithm~\ref{alg:parallel_sum}.
Note that in our pseudocode, for readability, we denote reads using the syntax:

\begin{algorithmic}
\Read{mods...}{args...} $f$(args...) \EndRead
\end{algorithmic}

\noindent This computation is static in the sense that regardless of updates, the RSP tree will always look the same. In general, our system supports arbitrary changes, even if they lead to wildly different trees.

\begin{algorithm}
  \caption{Parallel self-adjusting sum}
  \label{alg:parallel_sum}
  \footnotesize
  \begin{algorithmic}[1]
    \Function{Sum}{A[lo...hi] : \textbf{int} \algorithmicmod{} \textbf{array}, res : \textbf{int} \algorithmicmod{}}
      \If{lo = hi - 1}
        \Read{A[lo]}{x}
          \State \algwritemod{res}{x}
        \EndRead
      \Else
        \State \alglocal{} mid \algassign{} lo + (hi - lo) / 2
        \State \alglocal{} left\_res \algassign{} \algorithmicalloc{}(\textbf{int})
        \State \alglocal{} right\_res \algassign{} \algorithmicalloc{}(\textbf{int})
        \State \begin{varwidth}[t]{\linewidth}\algorithmicpar{}(\function{} $\Rightarrow$ \textsc{Sum}(A[lo...mid], left\_res), \\ \algtab
          \function{} $\Rightarrow$ \textsc{Sum}(A[mid...hi], right\_res))
        \end{varwidth}
        \Read{left\_res, right\_res}{x, y}
          \State \algwritemod{res}{x + y}
        \EndRead
      \EndIf
    \EndFunction
  \end{algorithmic}
\end{algorithm}

  \section{Change Propagation Algorithm}\label{sec:algorithms}

%\guy{I think we could use a more specific title than algorithms.
%  What kind of algorithm?}

We use a variant of SP trees (see introduction) extended with read (R) nodes, which we call RSP trees. A read is tracked in the RSP tree by creating an R node as the left child of the current S
node whenever a reader is executed. The reader code then executes in the subtree of the R node and the subsequent continuation proceeds in the sibling node. The full semantics for RSP trees for any given computation is defined by the algorithms in this section.
\begin{figure}
  \centering
  \includegraphics[width=0.4\columnwidth]{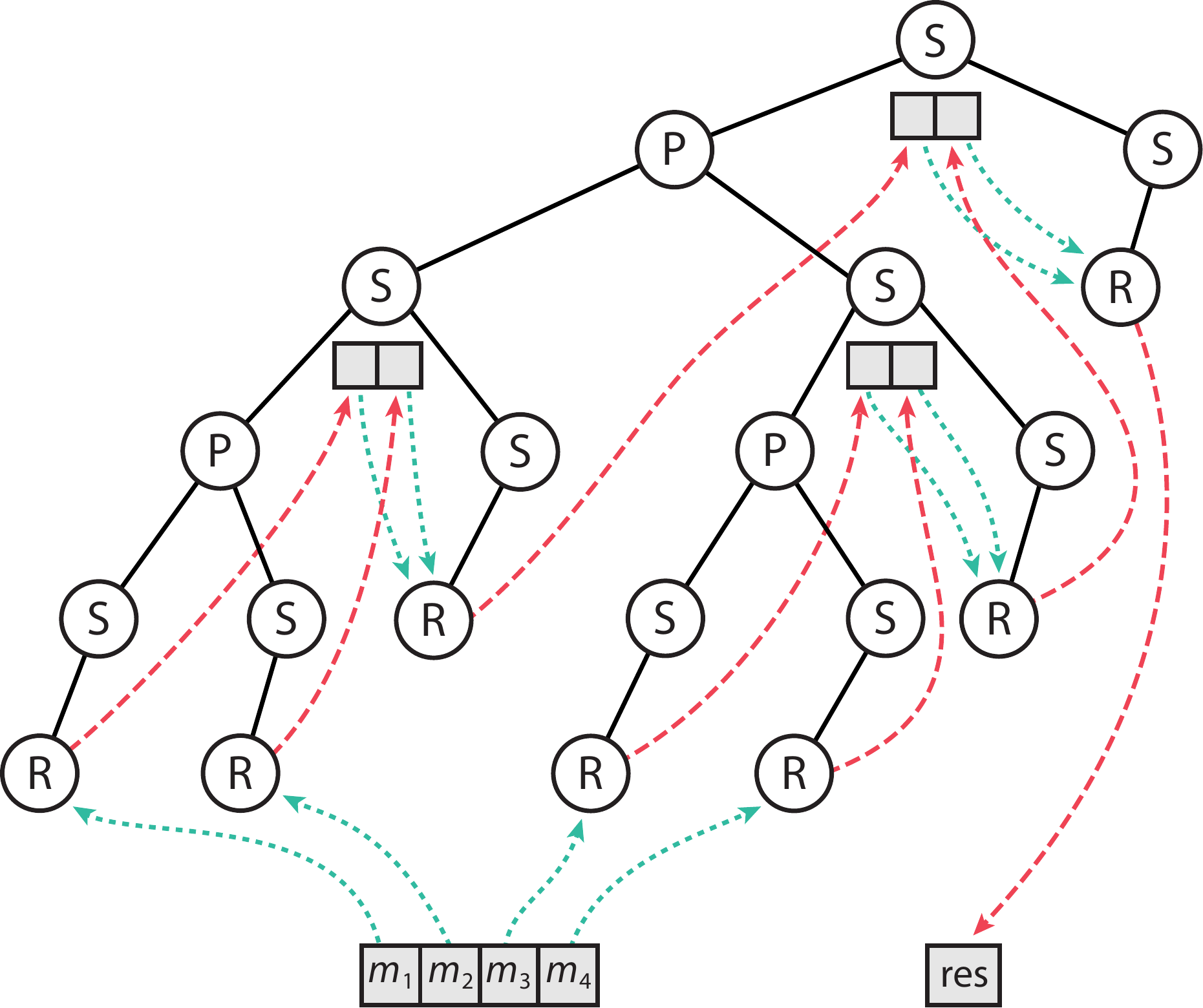}
  \caption{The RSP tree of the divide-and-conquer sum algorithm on an
    input of size four. Dynamically allocated modifiables are depicted
    underneath the $S$ node that allocated them. Writes and reads
    to/from modifiables are shown as red (long-dashed) and green
    (short-dashed) arrows respectively.}\label{fig:sp_tree}
\end{figure}

Figure~\ref{fig:sp_tree} shows an example RSP tree for the divide-and-conquer sum computation of Algorithm~\ref{alg:parallel_sum} on an input of size four. We use green (short-dashed) arrows to show which read nodes read from which modifiables, and red (long-dashed nodes) to indicate which sequential computations wrote to which modifiables. The four $R$ nodes lowest in the tree correspond to the reads of the input, which occur at the base case
of the algorithm. The $R$ nodes higher in the tree are the reads of the results of the
recursive calls.

Our framework facilitates self-adjusting computation by first building the RSP tree during the initial run of the static algorithm. To execute dynamic updates, when a modifiable is written to, all of the read nodes that read from it, and all of their ancestors in the RSP tree are marked as pending re-execution. Change propagation then simply consists in traversing the RSP tree, ignoring paths which are not marked, since no changes are present, and re-executing the marked readers. Note that this re-execution destroys the old portion of the RSP tree corresponding to the read and generates a new one, meaning that the old and new computations can be entirely structurally different. Additionally, such re-execution may also write to subsequent modifiables that were also read during the computation, so this process may mark additional nodes in the tree as pending re-execution, which will cause further propagation. The remainder
of this section discusses the high-level implementation of these operations and the framework's primitives.

Pseudocode for the key components of our algorithm is shown in Algorithms~\ref{alg:alloc_and_write}--\ref{alg:control}. In the code, \emph{current\_scope} is a thread-local variable pointing to the current S node of the RSP tree that the code is running in.

Maintaining this notion of scope is important for two reasons. Most obviously, it ensures that the RSP tree can be constructed while the algorithm is ran. Less obviously,
and more interestingly, it also allows us to more efficiently allocate and collect
dynamically allocated modifiables.

\myparagraph{Writing to modifiables}
Pseudocode for writing to modifiables is given in Algorithm~\ref{alg:alloc_and_write}. If the new value differs from the old value, all of the readers of that modifiable must be marked to trigger change propagation.

\begin{algorithm}
  \caption{Writing modifiables}
  \label{alg:alloc_and_write}
  \footnotesize
  \begin{algorithmic}[1]
    \Function{write}{dest : $\alpha\ \algorithmicmod{}$, value: $\alpha$}
    \If{dest is unwritten or value $\neq$ dest.val}
    \State dest.val \algassign{} value
    \ParallelFor{\textbf{each} reader in dest.readers}
      \State reader.affected \algassign{} \algtrue{}
      \State reader.mark()
    \EndParallelFor
    \EndIf
    \EndFunction
  \end{algorithmic}
\end{algorithm}

\myparagraph{Reading modifiables}
Pseudocode for the core read operation and for handling R nodes is shown in Algorithm~\ref{alg:reads}. R nodes consist of two specific fields, the list of modifiables that were read (mods), and the reader function that executes on the values of the modifiables (reader\_f). When an R node is created or destroyed, it adds or removes itself from the corresponding modifiables' list of readers respectively.
Note that this operation needs to happen atomically. We will discuss how we implement this in Section~\ref{sec:implementation}. Lastly, for added efficiency, we note that the size of the RSP tree can be reduced by creating the continuation S node (Line~\ref{line:read-continuation}) lazily, so that it will be omitted if there is no subsequent computation.

\begin{algorithm}
  \caption{Reading modifiables}
  \label{alg:reads}
  \footnotesize
  \begin{algorithmic}[1]
    \Function{read}{m : $(\alpha_1\ \algorithmicmod, ..., \alpha_k\ \algorithmicmod)$, r : $\alpha_1 \times ... \times \alpha_k \mapsto ()$}
    \State \alglocal cur \algassign{} current\_scope
    \State cur.left \algassign{} new $R$ node (m, r)
    \State current\_scope \algassign{} cur.left
    \State cur.left.\textproc{do\_read}()
    \State\label{line:read-continuation} cur.right \algassign{} new S node
    \State current\_scope \algassign{} cur.right
    \EndFunction
    
    \Function{$R$::create}{m, r}
    \State \algthis{}.mods \algassign{} m
    \State \algthis{}.reader\_f \algassign{} r
    \ParallelFor{\textbf{each} mod v \textbf{in} mods}
    \State\label{line:add_reader} v.readers \algassign{} v.readers $\cup \{ \algthis{} \}$  \algcomment{must be atomic!}
    \EndParallelFor
    \EndFunction
    
    \Function{$R$::do\_read}{}
    \State \alglocal{} $m_1, ..., m_k$ \algassign{} \algthis{}.mods
    \State \alglocal{} $v_1, ..., v_k$ \algassign{} $m_1$.val, ..., $m_k$.val
    \State \algthis{}.reader\_f($v_1, ..., v_k$)
    \EndFunction
    
    \Function{$R$::destroy}{}
    \For{\textbf{each} mod m \textbf{in} \algthis{}.mods}
    \State\label{line:remove_reader} m.readers $\algassign$ m.readers $\setminus$ $\{ \textbf{this} \}$ \algcomment{must be atomic!}
    \EndFor
    \EndFunction
  \end{algorithmic}
\end{algorithm}

\myparagraph{Parallelism}
The \algorithmicpar{} function creates a $P$ node as the left child of the current scope. The $P$ node has two $S$ nodes as children, which will correspond to the scope
of the two computations that run in parallel. After completing the
parallel computation, an $S$ node is created as the right child
of the current node to be the scope of any subsequent computation.
The algorithm is shown in Algorithm~\ref{alg:parallel}.

\begin{algorithm}
  \caption{Parallelism}
  \label{alg:parallel}
  \footnotesize
  \begin{algorithmic}[1]
    \Function{par}{left\_f: $() \mapsto ()$ , right\_f: $() \mapsto ()$}
    \State \alglocal cur \algassign{} current\_scope
    \State cur.left \algassign{} new $P$ node
    \State cur.left.left \algassign{} new $S$ node
    \State cur.left.right \algassign{} new $S$ node
    \State \textbf{in parallel do}:
    \State \hspace*{1em} \{ current\_scope \algassign{} cur.left.left; \ left\_f() \}
    \State \hspace*{1em} \{ current\_scope \algassign{} cur.left.right; \ right\_f() \}
    \State cur.right \algassign{} new $S$ node
    \State current\_scope \algassign{} cur.right
    \EndFunction
  \end{algorithmic}
\end{algorithm}

\myparagraph{Control operations}
These are depicted in Algorithm~\ref{alg:control}. Run creates the root S node of the RSP tree and runs the computation from scratch. The propagate functions perform change propagation for each of the kind of RSP tree nodes. Note that the P version propagates in parallel, and the S version sequentially. When reaching a read node, the algorithm reruns the associated reader function.

\begin{algorithm}
  \caption{Control operations}
  \label{alg:control}
  \footnotesize
  \begin{algorithmic}[1]
    \Function{run}{$f : () \mapsto ()$} : $S$
    \State \alglocal{} root \algassign{} new $S$ node
    \State current\_scope \algassign{} root
    \State $f$()
    \State \algreturn{} root
    \EndFunction
    
    \Function{propagate}{root : $S$}
    \If{root.marked} root.propagate() \EndIf
    \EndFunction
    
    \Function{node::mark}{}
    \State \algthis{}.marked \algassign{} \algtrue{}
    \If{\algthis{}.parent $\neq$ \algnull{} $\land\ \lnot$ \algthis{}.parent.marked}
    \State \algthis{}.parent.mark()
    \EndIf
    \EndFunction
    
    \Function{$S$::propagate}{}
    \If{\algthis{}.left $\neq$ \algnull{} $\land$ \algthis{}.left.marked}
    \State \algthis{}.left.propagate()
    \EndIf
    \If{\algthis{}.right $\neq$ \algnull{} $\land$ \algthis{}.right.marked}
    \State \algthis{}.right.propagate()
    \EndIf
    \State \algthis{}.marked \algassign{} \algfalse{}
    \EndFunction
    
    \Function{$P$::propagate}{}
    \If{\algthis{}.left.marked $\land$ \algthis{}.right.marked}
    \State \textbf{in parallel do}:
    \State \hspace*{1em} \algthis{}.left.propagate()
    \State \hspace*{1em} \algthis{}.right.propagate()
    \ElsIf{\algthis{}.left.marked}
     \algthis{}.left.propagate()
    \Else\ 
     \algthis{}.right.propagate()
    \EndIf
    \State \algthis{}.marked \algassign{} \algfalse{}
    \EndFunction
    
    \Function{$R$::propagate}{}
      \If{\algthis{}.affected}
        \State \algthis{}.left \algassign{} \algnull{}
        \State \algthis{}.right \algassign{} \algnull{}
        \State \algthis{}.\textproc{do\_read}()
        \State \algthis{}.affected \algassign{} \algfalse{}
      \Else
        \If{\algthis{}.left $\neq$ \algnull{} $\land$ \algthis{}.left.marked}
          \State \algthis{}.left.propagate()
        \EndIf
        \If{\algthis{}.right $\neq$ \algnull{} $\land$ \algthis{}.right.marked}
          \State \algthis{}.right.propagate()
        \EndIf
      \EndIf
      \State \algthis{}.marked \algassign{} \algfalse{}
    \EndFunction 
  \end{algorithmic}
\end{algorithm}

  \section{Analysis}

In this section, we provide an analysis of our model to establish its correctness
and prove bounds on the runtimes of our algorithms. 
Bounds in our analysis will depend on the work and span of the underlying algorithm, as well as the height of the generated RSP tree, which we note is at most the span of the algorithm, but can be much less. For all of our examples, it is at most $O(\log(n))$, even when the span of the algorithm is larger.

\myparagraph{Setting} For our analysis, we will consider algorithms $A$ in our parallel self-adjusting framework, which can be thought of as functions which act on given inputs $I = \{ (m,v) \}$, a set of modifiable-value pairs consisting of modifiables that $A$ will read, and their values. Executing $A(I)$ results in an output $(\tau, T)$, where state $\tau = \{ (m,v) \}$ denotes that for each $(m,v) \in \tau$, $A$ wrote the value $v$ to the modifiable $m$. $T$ is the RSP tree of the computation, where each read node is annotated with the reader function and the values that were read. We define the domain of a set of pairs $X$ by $dom(X) = \{ m : (m,v) \in X \}$. Due to the write-once restriction, note that in a valid execution, we must have $dom(I) \cap dom(\tau) = \emptyset$.
%A \emph{program state} $\sigma$ is a set of modifiable-value pairs that denotes the current values of all modifiables in a program. We can \emph{execute} $A$ on a program state $\sigma$ if there exists an $I \subseteq \sigma$ such that $dom(I)$ contains all modifiables that could be read by $A$.

We can then define a \emph{dynamic update} $\Delta = (I, I')$ to be a pair of input states with $I \neq I'$, denoting that the input is changed from $I$ to $I'$, which may involve changing the values of modifiables in $I$, adding new modifiables that were not read the first time, and removing modifiables that are no longer read. We can then think of change propagation as taking an RSP tree $T$ and a dynamic update $(I,I')$, and outputting a set of writes $\tau$ and an updated RSP tree $T'$.

We can now define the notion of \emph{affected readers}, which, intuitively, when applying an algorithm to two different inputs, are readers that exist in both versions of the computation but read different values, i.e.\ they are the frontiers at which the computations diverge.

\begin{definition}[Affected readers]
  Consider an algorithm $A$, two input states $I$ and $I'$, and their corresponding RSP trees $T$ and $T'$, i.e., $A(I) = (\tau, T)$ and $A(I') = (\tau', T')$ for some $\tau$ and $\tau'$. We say that a read node is \emph{subsumed} by another read node in the same tree if the first one is in the subtree of the second one. i.e., the first one was created while executing the second one's computation. Given two read nodes $v \in T$ and $v' \in T'$, we say that they are \emph{cognates} if the paths in $T$ and $T'$ to $v$ and $v'$ are the same, that is, the path branches left or right at the same time and have the same labels. We call a pair of cognate read nodes \emph{affected} if they read different values, and are not subsumed by another such node.
\end{definition}

\noindent Note that this definition of affected node makes sense because of the fact that computations in our framework are deterministic, and hence, the only place at which a computation can begin to differ is at a read node that reads different values than last time.
%Note that this definition does not even need to specifically mention whether the functions executed by the affected reader were the same, as they are guaranteed to be due to determinism. Read nodes that are subsumed by other affected read nodes are not considered affected because they may or may not even exist in the computation resulting from running the affected reader on the new values.
We now introduce the notion of \emph{computation distance}. The computation distance
models the amount of work required to re-execute the affected readers. In essence,
it is the minimum amount of work required to update the computation assuming absolutely no overhead.

\begin{definition}
\label{def:distance}
  Consider an algorithm $A$, two input states $I$ and $I'$, and their corresponding RSP trees $T$ and $T'$. Define the cost of a read node to be the work performed by its reader function\footnote{The work performed by the reader function is considered to be the work that it would perform when executed without self-adjusting computation, i.e., assuming that reads and writes take constant time.}. The computation distance between
  the executions of $A$ on the inputs $I$ and $I'$ is defined as the sum of the costs
  of the affected read nodes in $T$ and $T'$. More formally, if we denote by $l(T), v(T), w(T)$, the RSP label of the root node of $T$, the values read by the read node at the root of $T$, and the work performed by the reader function of the read node at the root of $T$, we can define the computation distance recursively as follows.
  \begin{equation}
  \begin{split}
        \delta(T, T') &= \begin{cases}
          w(T) + w(T') & \text{if } l(T) = R \land{} v(T) \neq v(T'),  \\
          \sum_{i=1}^k \delta(T_i, T_i') & \text{otherwise},
        \end{cases}
  \end{split}
  \end{equation} where $T_i$ denotes the $i^\text{th}$ subtree of $T$.
\end{definition}

\noindent Observe that due to determinism, the definition of computation distance will only consider cognate nodes $T, T'$ which must have the same number of children/subtrees. We are now ready to state the correctness theorem of our framework.

%Its proof and that of Theorem~\ref{thm:change-prop-fast} are provided in Appendix~\ref{appendix:proofs}.

\begin{theorem}[Correctness]\label{thm:change-prop-correct}
  Consider an algorithm $A$ and an input state $I$ where $A(I) = (\tau, T)$.
  %and a program state $\sigma$ such that $(I \cup \tau) \subseteq \sigma$.
  Let $\Delta = (I, I')$, where $A(I') = (\tau', T')$, denote a dynamic update to the input.
  %is applied to the program state $\sigma$.
  Then, applying change propagation to the RSP tree $T$ with dynamic update $\Delta$ yields
  \begin{enumerate}[leftmargin=*]
    \item writes $\tau''$ such that $\tau' \subseteq \tau'' \cup \{ (m,v) \in \tau\ |\ m \not\in dom(\tau'') \}$,
    \item the RSP tree $T'$.
  \end{enumerate}
\end{theorem}

\begin{proofsketch}
  Proving the correctness of change propagation essentially relies on establishing two facts: that it visits and re-executes all affected read nodes, and that re-executing just the affected read nodes is sufficient.
  
  The fact that all affected read nodes are re-executed can be established inductively on the sequential dependencies of the affected readers. The earliest affected reader must read a modifiable that exists in $I$ and $I'$ but has a different value, and hence will be marked in the RSP tree and will be re-executed. An affected reader that has had all of its sequential dependencies re-executed must be marked since it either reads a modifiable that exists in $I$ and $I'$ but has a different value, or it reads a modifiable that is written earlier in the computation. In the second case, since computations are deterministic, the modifiable must be written inside a reader whose input has changed, and hence is an affected reader which has already been re-executed.
  
  Establishing that re-executing all affected readers writes to all modifiables whose values in $\tau'$ are different than in $\tau$ follows from determinism and the write-once restriction. Determinism implies that all differing writes must occur inside an affected reader, and the write-once restriction ensures that these writes exist in $\tau'$. Lastly, the fact that the RSP tree is updated to $T'$ follows from determinism.
\end{proofsketch}

\begin{proofsketch}[Proof sketch of Theorem~\ref{thm:change-prop-fast}]
  Since there are $O(R_\Delta)$ affected readers, it costs at most $O(R_\Delta \cdot h)$ work to traverse the RSP tree to reach each of them. The work required to re-execute all affected readers and destroy their old RSP subtrees is $O(W_\Delta)$ by definition, plus any overhead encountered from maintaining modifiables' reader sets and marking ancestors when performing the \algorithmicwrite{} primitive. We can argue that these overheads can be reduced to constant or amortized. To reduce the maintenance of reader sets to constant overhead, the algorithm can maintain each modifiable's reader set as a hashtable. To avoid issues of concurrency and resizing, insertions and deletions (Lines~\ref{line:add_reader}~and~\ref{line:remove_reader} in Algorithm~\ref{alg:reads}) can be deferred and performed in batch after change propagation is complete. The overhead of \algorithmicwrite{} can be amortized by noticing that for all marked nodes, they will either be traversed by change propagation, or destroyed by a re-execution.
  
  Lastly, we consider how these overheads affect the span of the algorithm. Each \algorithmicwrite{} operation takes up to $h$ time, and each \algorithmicread{} may require a hashtable operation that takes up to $\log(r)$ time w.h.p, where $r$ is the size of the reader set. However, $h \geq \log(r)$ and hence the overhead is at most $h$ per operation w.h.p., leading to a total span of $O(s \cdot h)$ w.h.p.
\end{proofsketch}

\noindent It is worth noting that the randomness in our bounds comes purely from the use of hashtables to store the reader sets. For algorithms in which each modifiable has only a constant number of readers, which is very often the case, the bounds can therefore be made deterministic.
%The remainder of this section discusses the context of these results and some additional observations about the overhead of the framework.

\myparagraph{Analyzing the computation distance of algorithms}
To obtain bounds for dynamic updates on particular algorithms implemented in our framework, it suffices to analyze the number of affected reads and the computation distance for the desired class of updates (and the span of the algorithm which is usually already known). Here, we will sketch an analysis of the sum algorithm from Algorithm~\ref{alg:parallel_sum}.

\begin{theorem}
Consider Algorithm~\ref{alg:parallel_sum} on an input $A$ of $n$ modifiables, and a dynamic update in which the values of $k$ modifiables are changed. The number of affected reads and the computation distance induced by such an update is $O(k\log(1+n/k))$.
\end{theorem}

\begin{proofsketch}
Note that the algorithm performs $\log(n)$ levels of recursion. We count separately the number of affected reads that occur during the first $\log(k)$ levels and those that occur after. During the first $\log(k)$ levels, since the algorithm performs binary recursion, there can be no more than $O(2^{\log(k)}) = O(k)$ reads in total, affected or not. The $k$ updated modifiables will affect $k$ of the base-case reads on Line~3. The corresponding writes on Line~4 then affect up to $k$ reads on Line~10 from the calling functions. The writes on Line~11 then affect up to $k$ reads from their callers, and so on. The final $\log(n/k)$ levels of recursion therefore account for at most $k\log(n/k)$ affected readers. Therefore, in total, there can be at most $O(k + k\log(n/k)) = O(k \log(1+n/k))$ affected readers, each of which performs $O(1)$ work.
\end{proofsketch}

\noindent In ~\cite{acar2020batch}, several algorithms, including list contraction and tree contraction, which also appear in our benchmarks, had their computation distance analyzed in the round-synchronous model. The round-synchronous model can be implemented in our framework, and hence it is straightforward to translate these analyses to bounds our framework.
%In the full version of the paper, we will provide more example analyses of computation distance.

\myparagraph{Overhead of self-adjusting computation} In addition to the cost of dynamic updates, we can also discuss the overhead of the initial computation. Note that each node in the RSP tree corresponds to at least one primitive operation, and hence the cost of the building the tree and later destroying it can be charged to the computation. Then, just as in change propagation, the overhead of the \algorithmicread{} and \algorithmicwrite{} primitives are either constant, or can be amortized (see the Proof sketch of Theorem~\ref{thm:change-prop-fast}), leading to constant amortized overhead.

Lastly, we remark on the memory usage. The two sources of memory overhead come from the RSP tree and modifiables. In the worst case, the size of the RSP tree is proportional to the work of the algorithm. However, for any sensible algorithm, both strands of any parallel fork will contain at least one read (if they do not, the parallel fork was unnecessary). Therefore, under this assumption, the size of the SP tree is proportional to the number of reads in the algorithm. Since the memory overhead of modifiables (their reader sets) is also proportional to the number of reads, the additional memory overhead is just proportional to the number of reads.

\myparagraph{Work-efficiency of change propagation}
By definition, re-executing a set of affected readers of computation distance $W_\Delta$ must take $O(W_\Delta)$ work. Based on Theorem~\ref{thm:change-prop-fast}, we therefore consider the work overhead of change propagation to be $R_\Delta \cdot h$. This means that if $W_\Delta \geq R_\Delta \cdot h$, i.e, each affected reader performs at least $h$ work on average, then change propagation essentially has just constant-time overhead. In practice, this suggests that good granularity control is important for writing efficient self-adjusting algorithms.

\myparagraph{Comparison to sequential self-adjusting computation}
The best sequential algorithms for self-adjusting computation~\cite{acar2004dynamizing} can propagate an update of computation distance $W_\Delta$ in $O(W_\Delta \log(W_\Delta))$ work. Compared to our bounds, which are at most $O(W_\Delta \cdot h)$, the difference is a $\log(W_\Delta)$ versus $h$. Given a parallel algorithm on input size $n$ with $\text{polylog}(n)$ span, we have $h \leq \text{polylog}(n)$. However, often, and for every example we studied, $h$ is just $\log(n)$, even for algorithms with larger span. Therefore at worst, our algorithm is $O(\text{polylog}(n))$ slower than the best sequential algorithm, but in the common case, just $O(\log(n)/\log(W_\Delta))$ slower.

  \section{Implementation}\label{sec:implementation}

To study its practical performance, we implemented our framework as a library for \cpp{}.
%\footnote{Our code is publicly available at \url{https://github.com/cmuparlay/psac}}
For parallelism, we use the work-stealing scheduler from the Parlay library~\cite{blelloch2020parlaylib}. For memory allocation, we use jemalloc~\cite{evans2006scalable} in addition to Parlay's pool-based memory allocator. In this section, we discuss some of the interesting aspects of the implementation of the system, and note some useful optimizations.
%We then finish with a simple example program.

\myparagraph{Reader set implementation}
One interesting part of the system is handling the reader sets of
modifiables. Since multiple concurrently executing threads may read the same
modifiable, it is important that modifications to this set are thread-safe.
To obtain our theoretical bounds, we describe the algorithm
using a hashtable. In practice, however, we observe that the majority
of modifiables in self-adjusting algorithms have just a small constant
number of readers, often just one. We therefore implement the reader sets
with a hybrid data structure that stores a single reader inline with no
heap allocation when there is only one reader. When the number of readers
becomes more than one, the reader set atomically converts itself into
a linked data structure. We used a linked list for algorithms with small
reader sets, and a randomized binary search tree for algorithms with larger
reader sets.

Our binary search tree uses the hashes of the addresses of the
reader nodes as the random keys. To insert a new reader into the tree, our
algorithms attempts to insert it into the appropriate leaf of the tree using
an atomic compare-and-swap (CAS) operation. If the CAS succeeds, the insertion
is successful. Otherwise, note that the correct location for the key must be a
child of the node that instead won the CAS, so our algorithm proceeds down and
tries again. To simplify deletions, rather than deleting from the tree eagerly,
nodes that need to be removed are simply marked as dead, and removed during the
next traversal.

To ensure thread safety, we have to make sure that operations on the reader
sets that might race are safe. Note that insertions correspond to reads,
traversals correspond to writes, and deletions correspond to the cleanup
of destroyed subtrees after a computation is re-executed. Insertions will
therefore never race with traversals since reads and writes to the same
modifiable can not race in a valid self-adjusting program. Deletions may
however race with traversals or insertions since the cleanup of an RSP subtree
may take place while another re-computation is occurring. One way to mitigate
any potential problems is to defer all destructions of RSP subtrees until a later
garbage-collection phase, rather than performing them during change propagation.
Lastly, multiple traversals can not race due to the write-once condition,
but multiple insertions or deletions can. Our algorithm is safe with respect to
concurrent insertions, and our lazy deletion strategy makes concurrent
deletions safe.

\myparagraph{Garbage collection}
Rather than eagerly deleting subtrees of the RSP tree when a reader is re-executed,
we instead move such subtrees off to a garbage pile which we collect  after performing
change propagation. This simplifies the
destruction of subtrees since doing so naively can very easily lead to race
conditions. For example, if performing eager deletion, it is possible that while
a subtree is being collected, a concurrent computation causes one of its $R$ nodes
to be marked, which could lead to marking a node that has already been deleted.
%Rather than handling this with a more expensive option such as reference counting,
%we instead decided to perform delayed garbage collection.
%, since this also improves the
%responsiveness of change propagation, as the result of the algorithm can be made
%visible before the garbage collector is ran.

\myparagraph{Supporting dynamically sized inputs} Modifiables give us the ability to easily write algorithms that support updating the \emph{values} in the input and propagating the results. In many situations, we also want to support the ability to add/remove elements to/from the input. In sequential self-adjusting computation, this is achieved by using linked lists to represent the input. In the parallel setting, we can achieve similar results by representing the input as a balanced binary tree. The trick is to use modifiables to represent the parent/children relationships in the tree so that if a new element is inserted, this will cause an update of a child modifiable and allow change propagation to update the computation with the new element.

  \section{Benchmarks and Evaluation}

In this section, we evaluate the practical performance of our
system. We implemented six benchmarks, exhibiting a range of different
characteristics and providing different insights into the
quality of the proposed algorithms.

\myparagraph{Experimental setup} We ran our experiments on a
4-socket AMD machine with 32 physical cores in total, each
running at 2.4 GHz, with 2-way hyperthreading, a 6MB L3
cache per socket, and 200 GB of main memory. All of our code
was compiled using Clang 9 with optimization level \texttt{-O3}.
Each experiment
was run using 1 -- 64 worker threads in increasing powers of
two. We used the Google Benchmark C++ library to measure the
speed (in real/wall time) of each benchmark.
We run each benchmark ten times and take the average running time.

\myparagraph{Benchmark setup} Each benchmark consists
of four parts. First, we run a static sequential program
and a static parallel program that implement the same algorithm
to the self-adjusting one but without any
overhead from self-adjusting computation.
We then benchmark the parallel self-adjusting program, both on
its initial computation, and on performing dynamic updates
with change propagation. For each of the examples, we
use varying batch update sizes to measure the effect that
batch sizes have on the amount of parallelism exhibited by the
update, and the amount of work required to propagate it. We do not
include the time taken to perform garbage collection in the
measurements.

We hoped to experimentally compare our results to those of \cite{burckhardtlesayiba11}, but their code is not publicly available, and we were unable, despite our efforts, to obtain it.

\myparagraph{Reporting of results}
For each benchmark, we provide numerical results in Tables~\ref{tab:edit-distance-results}--\ref{tab:filter-results},
which show the running times of the static sequential
algorithm (Seq), parallel static algorithm (Parallel Static),
the initial computation of the self-adjusting algorithm
(PSAC Compute), and the dynamic
updates (PSAC Update), for $1$ processor (1), $32$ processors (32), and
$32$ processors with hyperthreading (32ht). For
each of the parallel algorithms, we compute
the self-speedup (SU), which is the relative improvement of the
32 or 32ht performance (whichever is better) compared to the
1 processor performance. For each example, we measure the performance
for some fixed input size $n$ and varying batch update sizes $k$.
%Note that
%for the static algorithms, their runtime is independent of the
%batch update size, since they simply recompute from scratch.

For the dynamic updates, we measure work savings (WS), which is the relative improvement
of their 1 processor performance
compared to the static sequential algorithm. Finally, we report
the total speedup, which is the relative performance of the dynamic updates with 32 or 32ht processors
compared to the static sequential algorithm (equivalently, the
product of the speedup and the work savings). This allows us
to measure separately, the benefits due to parallelism (the SU),
the benefits due to dynamism (the WS), and their total combined benefit (Total).

\myparagraph{Applications}
We implemented the following benchmarks.
%We provide thorough descriptions of each in Appendix~\ref{appendix:benchmarks}.

\begin{itemize}[leftmargin=*]
\item \textbf{Spellcheck:} Computes the minimum edit distance of a set of one million strings to a target string.
\item \textbf{Raytracer:} Renders a $2000 \times 2000$ pixel scene consisting of three reflective balls using a simple ray tracing method.
\item \textbf{String Hash:} Computes the Rabin-Karp fingerprint (hash) of a one-hundred-million character string.
\item \textbf{Dynamic Sequence:} Computes a list contraction of a linked list of length one million.
\item \textbf{Dynamic Trees:} Computes a tree contraction of a tree on one million nodes.
\item \textbf{Filter:} Filters the elements of a BST with respect to a given predicate, returning a new BST.
\end{itemize}

\subsection{Results}

The results of our experiments are depicted in Tables~\ref{tab:edit-distance-results}--\ref{tab:filter-results}.

\begingroup
\renewcommand*{\arraystretch}{1.1}

\begin{table}[p]
\footnotesize
\centering
\begin{subfigure}{0.99\columnwidth}
\centering
\noindent\begin{tabularx}{0.95\columnwidth}{| Y | Y | Y | Y | Y | Y | Y|} 
\hline \multicolumn{7}{|c|}{\textbf{Static Algorithm}} \\ 
\hline $\mathbf{n}$ & $\mathbf{1t}$ & $\mathbf{32}$\textbf{t} & $\mathbf{32}$\textbf{ht} & \textbf{SU} & \multicolumn{2}{|c|}{\textbf{Seq Baseline}} \\ 
\hline $10^6$ & 36.21s & 1.19s & 908ms & 39.84 & \multicolumn{2}{|c|}{36.72s} \\ 
\hline \multicolumn{7}{|c|}{\textbf{PSAC Initial Run}} \\ 
\hline $\mathbf{n}$ & $\mathbf{1t}$ & $\mathbf{32}$\textbf{t} & $\mathbf{32}$\textbf{ht} & \textbf{SU} & \multicolumn{2}{|c|}{} \\ 
\hline $10^6$ & 36.89s & 1.20s & 750ms & 49.16 & \multicolumn{2}{|c|}{} \\\hline \multicolumn{7}{|c|}{\textbf{PSAC Dynamic Update}} \\ 
\hline $\mathbf{k}$ & $\mathbf{1t}$ & $\mathbf{32}$\textbf{t} & $\mathbf{32}$\textbf{ht} & \textbf{SU} & \textbf{WS} & \textbf{T} \\ 
\hline $10^0$ & 44us & 45us & 47us & 0.98 & 819.6k & 800.4k \\ 
\hline $10^1$ & 445us & 90us & 112us & 4.93 & 82.42k & 406.7k \\ 
\hline $10^2$ & 4.36ms & 265us & 208us & 20.92 & 8.41k & 176.0k \\ 
\hline $10^3$ & 44ms & 2.03ms & 1.20ms & 36.97 & 827.9 & 30.61k \\ 
\hline $10^4$ & 436ms & 19ms & 11ms & 37.46 & 84.18 & 3.15k \\ 
\hline $10^5$ & 4.04s & 167ms & 118ms & 34.08 & 9.10 & 310.2 \\ 
\hline $10^6$ & 37.67s & 1.81s & 1.21s & 31.04 & 0.97 & 30.26 \\ 
\hline 
\end{tabularx}
\end{subfigure}
%\begin{subfigure}{0.70\columnwidth}
%  \centering
%  \includegraphics[width=\textwidth]{plots/edit_distance-1000000-abs-update-throughput}
%\end{subfigure}
\medskip
\caption{Benchmark results for Spellcheck.}\label{tab:edit-distance-results}
\end{table}

\begin{table}[p]
\footnotesize
\centering
\begin{subfigure}{0.99\columnwidth}
\centering
\noindent\begin{tabularx}{0.95\columnwidth}{| Y | Y | Y | Y | Y | Y | Y|} 
\hline \multicolumn{7}{|c|}{\textbf{Static Algorithm}} \\ 
\hline $\mathbf{n}$ & $\mathbf{1t}$ & $\mathbf{32}$\textbf{t} & $\mathbf{32}$\textbf{ht} & \textbf{SU} & \multicolumn{2}{|c|}{\textbf{Seq Baseline}} \\ 
\hline - & 2.46s & 80ms & 49ms & 49.32 & \multicolumn{2}{|c|}{2.54s} \\ 
\hline \multicolumn{7}{|c|}{\textbf{PSAC Initial Run}} \\ 
\hline $\mathbf{n}$ & $\mathbf{1t}$ & $\mathbf{32}$\textbf{t} & $\mathbf{32}$\textbf{ht} & \textbf{SU} & \multicolumn{2}{|c|}{} \\ 
\hline - & 13.65s & 520ms & 323ms & 42.24 & \multicolumn{2}{|c|}{} \\\hline \multicolumn{7}{|c|}{\textbf{PSAC Dynamic Update}} \\ 
\hline $\mathbf{k}$ & $\mathbf{1t}$ & $\mathbf{32}$\textbf{t} & $\mathbf{32}$\textbf{ht} & \textbf{SU} & \textbf{WS} & \textbf{T} \\ 
\hline - & 112ms & 8.66ms & 7.47ms & 15.08 & 26.44 & 398.8 \\ 
\hline 
\end{tabularx}
\end{subfigure}
%\begin{subfigure}{0.70\columnwidth}
%  \centering
%  \includegraphics[width=\textwidth]{plots/raytrace-1-rel-scaling}
%\end{subfigure}
\medskip
\caption{Benchmark results for Raytracer.}\label{tab:raytracer-results}
\end{table}

\begin{table}[p]
\footnotesize
\centering
\begin{subfigure}{0.99\columnwidth}
\centering
\noindent\begin{tabularx}{0.95\columnwidth}{| Y | Y | Y | Y | Y | Y | Y|} 
\hline \multicolumn{7}{|c|}{\textbf{Static Algorithm}} \\ 
\hline $\mathbf{n}$ & $\mathbf{1t}$ & $\mathbf{32}$\textbf{t} & $\mathbf{32}$\textbf{ht} & \textbf{SU} & \multicolumn{2}{|c|}{\textbf{Seq Baseline}} \\ 
\hline $10^8$ & 1.86s & 58ms & 31ms & 58.34 & \multicolumn{2}{|c|}{1.72s} \\ 
\hline \multicolumn{7}{|c|}{\textbf{PSAC Initial Run}} \\ 
\hline $\mathbf{n}$ & $\mathbf{1t}$ & $\mathbf{32}$\textbf{t} & $\mathbf{32}$\textbf{ht} & \textbf{SU} & \multicolumn{2}{|c|}{} \\ 
\hline $10^8$ & 3.05s & 96ms & 61ms & 49.81 & \multicolumn{2}{|c|}{} \\\hline \multicolumn{7}{|c|}{\textbf{PSAC Dynamic Update}} \\ 
\hline $\mathbf{k}$ & $\mathbf{1t}$ & $\mathbf{32}$\textbf{t} & $\mathbf{32}$\textbf{ht} & \textbf{SU} & \textbf{WS} & \textbf{T} \\ 
\hline $10^0$ & 14us & 16us & 16us & 0.90 & 118.8k & 107.1k \\ 
\hline $10^1$ & 134us & 66us & 74us & 2.02 & 12.81k & 25.86k \\ 
\hline $10^2$ & 1.26ms & 162us & 122us & 10.35 & 1.36k & 14.10k \\ 
\hline $10^3$ & 12ms & 717us & 512us & 25.17 & 133.3 & 3.36k \\ 
\hline $10^4$ & 103ms & 4.63ms & 3.11ms & 33.16 & 16.68 & 553.0 \\ 
\hline $10^5$ & 621ms & 26ms & 18ms & 34.20 & 2.77 & 94.76 \\ 
\hline $10^6$ & 2.49s & 108ms & 66ms & 37.27 & 0.69 & 25.73 \\ 
\hline $10^7$ & 5.21s & 213ms & 132ms & 39.30 & 0.33 & 12.99 \\ 
\hline $10^8$ & 19.47s & 709ms & 472ms & 41.24 & 0.09 & 3.65 \\ 
\hline 
\end{tabularx}
\end{subfigure}
%\begin{subfigure}{0.70\columnwidth}
%  \centering
%  \includegraphics[width=\textwidth]{plots/rabin_karp-100000000-abs-update-throughput}
%\end{subfigure}
\medskip
\caption{Benchmark results for String Hash.}\label{tab:rabin-karp-results}
\end{table}

\begin{table}[p]
\footnotesize
\centering
\begin{subfigure}{0.99\columnwidth}
\centering
\noindent\begin{tabularx}{0.95\columnwidth}{| Y | Y | Y | Y | Y | Y | Y|} 
\hline \multicolumn{7}{|c|}{\textbf{Static Algorithm}} \\ 
\hline $\mathbf{n}$ & $\mathbf{1t}$ & $\mathbf{32}$\textbf{t} & $\mathbf{32}$\textbf{ht} & \textbf{SU} & \multicolumn{2}{|c|}{\textbf{Seq Baseline}} \\ 
\hline $10^6$ & 647ms & 70ms & 73ms & 8.77 & \multicolumn{2}{|c|}{586ms} \\ 
\hline \multicolumn{7}{|c|}{\textbf{PSAC Initial Run}} \\ 
\hline $\mathbf{n}$ & $\mathbf{1t}$ & $\mathbf{32}$\textbf{t} & $\mathbf{32}$\textbf{ht} & \textbf{SU} & \multicolumn{2}{|c|}{} \\ 
\hline $10^6$ & 4.32s & 219ms & 464ms & 19.66 & \multicolumn{2}{|c|}{} \\\hline \multicolumn{7}{|c|}{\textbf{PSAC Dynamic Update}} \\ 
\hline $\mathbf{k}$ & $\mathbf{1t}$ & $\mathbf{32}$\textbf{t} & $\mathbf{32}$\textbf{ht} & \textbf{SU} & \textbf{WS} & \textbf{T} \\ 
\hline $10^0$ & 761us & 629us & 736us & 1.21 & 770.2 & 931.0 \\ 
\hline $10^1$ & 5.58ms & 1.65ms & 1.97ms & 3.38 & 105.2 & 355.2 \\ 
\hline $10^2$ & 31ms & 3.31ms & 2.95ms & 10.69 & 18.62 & 199.0 \\ 
\hline $10^3$ & 201ms & 14ms & 10ms & 18.53 & 2.91 & 53.94 \\ 
\hline $10^4$ & 1.26s & 74ms & 53ms & 23.33 & 0.47 & 10.89 \\ 
\hline $10^5$ & 5.11s & 263ms & 195ms & 26.15 & 0.11 & 3.00 \\ 
\hline $10^6$ & 8.45s & 624ms & 492ms & 17.14 & 0.07 & 1.19 \\ 
\hline 
\end{tabularx}
\end{subfigure}
%\begin{subfigure}{0.70\columnwidth}
%  \centering
%  \includegraphics[width=\textwidth]{plots/list_contraction-1000000-abs-update-throughput}
%\end{subfigure}
\medskip
\caption{Benchmark results for Dynamic Sequence.}\label{tab:list-contraction-results}
\end{table}

\begin{table}[p]
\footnotesize
\centering
\begin{subfigure}{0.99\columnwidth}
\centering
\noindent\begin{tabularx}{0.95\columnwidth}{| Y | Y | Y | Y | Y | Y | Y|} 
\hline \multicolumn{7}{|c|}{\textbf{Static Algorithm}} \\ 
\hline $\mathbf{n}$ & $\mathbf{1t}$ & $\mathbf{32}$\textbf{t} & $\mathbf{32}$\textbf{ht} & \textbf{SU} & \multicolumn{2}{|c|}{\textbf{Seq Baseline}} \\ 
\hline $10^6$ & 915ms & 85ms & 66ms & 13.85 & \multicolumn{2}{|c|}{824ms} \\ 
\hline \multicolumn{7}{|c|}{\textbf{PSAC Initial Run}} \\ 
\hline $\mathbf{n}$ & $\mathbf{1t}$ & $\mathbf{32}$\textbf{t} & $\mathbf{32}$\textbf{ht} & \textbf{SU} & \multicolumn{2}{|c|}{} \\ 
\hline $10^6$ & 4.85s & 242ms & 689ms & 20.02 & \multicolumn{2}{|c|}{} \\\hline \multicolumn{7}{|c|}{\textbf{PSAC Dynamic Update}} \\ 
\hline $\mathbf{k}$ & $\mathbf{1t}$ & $\mathbf{32}$\textbf{t} & $\mathbf{32}$\textbf{ht} & \textbf{SU} & \textbf{WS} & \textbf{T} \\ 
\hline $10^0$ & 698us & 584us & 672us & 1.19 & 1.18k & 1.41k \\ 
\hline $10^1$ & 3.28ms & 1.04ms & 1.23ms & 3.14 & 251.7 & 789.4 \\ 
\hline $10^2$ & 24ms & 2.29ms & 2.18ms & 11.03 & 34.23 & 377.7 \\ 
\hline $10^3$ & 210ms & 12ms & 10ms & 20.46 & 3.93 & 80.33 \\ 
\hline $10^4$ & 1.47s & 79ms & 60ms & 24.33 & 0.56 & 13.68 \\ 
\hline $10^5$ & 5.19s & 254ms & 173ms & 29.85 & 0.16 & 4.74 \\ 
\hline $10^6$ & 8.59s & 428ms & 306ms & 28.00 & 0.10 & 2.69 \\ 
\hline 
\end{tabularx}
\end{subfigure}
%\begin{subfigure}{0.70\columnwidth}
%  \centering
%  \includegraphics[width=\textwidth]{plots/tree_contraction-1000000-abs-update-throughput}
%\end{subfigure}
\medskip
\caption{Benchmark results for Dynamic Trees.}\label{tab:tree-contraction-results}
\end{table}

\begin{table}[p]
\footnotesize
\centering
\begin{subfigure}{0.99\columnwidth}
\centering
\noindent\begin{tabularx}{0.95\columnwidth}{| Y | Y | Y | Y | Y | Y | Y|} 
\hline \multicolumn{7}{|c|}{\textbf{Static Algorithm}} \\ 
\hline $\mathbf{n}$ & $\mathbf{1t}$ & $\mathbf{32}$\textbf{t} & $\mathbf{32}$\textbf{ht} & \textbf{SU} & \multicolumn{2}{|c|}{\textbf{Seq Baseline}} \\ 
\hline $10^7$ & 361ms & 17ms & 15ms & 23.09 & \multicolumn{2}{|c|}{262ms} \\ 
\hline \multicolumn{7}{|c|}{\textbf{PSAC Initial Run}} \\ 
\hline $\mathbf{n}$ & $\mathbf{1t}$ & $\mathbf{32}$\textbf{t} & $\mathbf{32}$\textbf{ht} & \textbf{SU} & \multicolumn{2}{|c|}{} \\ 
\hline $10^7$ & 630ms & 35ms & 31ms & 20.27 & \multicolumn{2}{|c|}{} \\\hline \multicolumn{7}{|c|}{\textbf{PSAC Dynamic Update}} \\ 
\hline $\mathbf{k}$ & $\mathbf{1t}$ & $\mathbf{32}$\textbf{t} & $\mathbf{32}$\textbf{ht} & \textbf{SU} & \textbf{WS} & \textbf{T} \\ 
\hline $10^0$ & 36us & 109us & 128us & 0.33 & 13.20k & 4.36k \\ 
\hline $10^1$ & 275us & 242us & 298us & 1.14 & 1.73k & 1.96k \\ 
\hline $10^2$ & 2.74ms & 1.39ms & 1.25ms & 2.19 & 173.6 & 380.9 \\ 
\hline $10^3$ & 25ms & 3.73ms & 3.69ms & 6.95 & 18.56 & 129.1 \\ 
\hline $10^4$ & 143ms & 9.96ms & 8.18ms & 17.50 & 3.32 & 58.19 \\ 
\hline $10^5$ & 543ms & 31ms & 23ms & 22.97 & 0.88 & 20.13 \\ 
\hline $10^6$ & 1.04s & 80ms & 55ms & 18.76 & 0.46 & 8.61 \\ 
\hline $10^7$ & 2.12s & 198ms & 159ms & 13.28 & 0.22 & 2.98 \\ 
\hline 
\end{tabularx}
\end{subfigure}
%\begin{subfigure}{0.70\columnwidth}
%  \centering
%  \includegraphics[width=\textwidth]{plots/filter-10000000-abs-update-throughput}
%\end{subfigure}
\medskip
\caption{Benchmark results for Filter.}\label{tab:filter-results}
\end{table}

\endgroup

\myparagraph{The initial run}
We are interested in the overhead of the initial run. This is the ratio of the runtime of the self-adjusting algorithm compared to the sequential baseline. Prior work on sequential self-adjusting computation~\cite{acar2009experimental} observed overheads ranging from 1.9 to 29 depending on the application.

The overhead of the initial run varies with the problem and the granularity of the work performed by the readers. For the spellcheck benchmark, the overhead is negligible since each of the readers performs a relatively expensive edit distance computation, completely hiding the overhead of the framework. For algorithms with smaller granularity, such as the Rabin-Karp benchmark, we observe work overheads of around 1.7. The filter algorithm also uses a similar granularity, and hence experiences similarly low overhead.

On the other hand, the raytracing algorithm involves modifiables with a large number of readers, so the work overhead is higher, at around a factor of 4.6. The list contraction and tree contraction benchmarks both perform $O(\log(n))$ rounds of computation, with dependency chains spanning across them, and hence have larger overheads of 5.8 and 7.3.

\myparagraph{Work savings}
Work savings measure the relative improvement in runtime from using self-adjusting computation to perform an update compared to running the algorithm from scratch. As was the case for the work overhead, the work savings are dependent on the granularity of the work performed by the readers. Of course, the work savings are also heavily dependent on the size of the update relative to the size of the entire input. For small updates, the work savings range from 770 when updating one element of one million elements (list contraction) to 819k when updating one of one million strings (edit distance). The work savings for the raytracer benchmark are also very encouraging. For the given dynamic update, a total of $6.25\%$ of the image needed to be updated, but the change propagation algorithm performed approximately $4\%$ of the work required to recompute from scratch. For benchmarks with varying update sizes, the work savings gradually decrease as the update size increases. It is interesting to look at the crossover point where from-scratch execution becomes more efficient than change propagation.

For benchmarks like spellcheck, which perform heavy work at the reads, from-scratch execution does not outperform change propagation until updating the entire input. For modest-granularity benchmarks like hashing, from-scratch execution wins when the update size reaches $k = 10^6$ for sequential execution, or $k = 10^5$ for parallel execution, on an initial input of size $n = 10^8$. For tree contraction and list contraction, the crossover points occur at $k = 10^4$ out of $n = 10^6$ elements for both sequential and parallel execution. For filter, the crossover point occurs at roughly $k = 10^5$ out of $n = 10^7$ elements. In general, crossovers tend to occur a couple of orders of magnitude before the input size.

\myparagraph{Speedup}
The initial runs of our algorithms all benefit, often substantially, from parallelism. On 32 hyperthreaded cores (64 threads), spellcheck and hashing experience parallel speedups of 49-50x. Raytracing achieves 42x, and list contraction, tree contraction, and filter speed up 19-20x.

In addition to the initial run, updates also benefit from parallelism, particularly as the update sizes increases. Although there is little potential for parallelism for $k = 1$ updates, each benchmark exhibits speedups ranging from 22-39x for larger update sizes. At the crossover points, where change propagation is still competitive with from-scratch execution, speedups range from 22-34x. This further supports the notion that parallelism and self-adjusting computation are highly complementary methods. Self-adjusting computation leads to substantial savings for small update sizes, and parallelism provides strong speedups for larger update sizes. For moderate update sizes, both are effective and their benefits combine to yield good total performance improvements.

\myparagraph{Tree size and memory usage}
The RSP tree size, and hence the memory overhead of a self-adjusting algorithm depends heavily on
the \emph{granularity} at which the data is stored and processed. Table~\ref{tab:memory} shows the RSP
tree sizes and memory usage of each of our benchmarks at their default granularity. For most algorithms, the memory overhead ranges between 1-7x the
input size, which is consistent with prior work on sequential self-adjusting computation~\cite{acar2009experimental}.
The outliers are our list contraction and tree contraction benchmarks, which use substantially more memory because
they perform $O(n\log(n))$ work over $O(\log(n))$ rounds of computation, all of which is represented
in the RSP tree, essentially leading to the tree size being an additional factor of $\log(n)$ larger than the input.
A more sophisticated implementation of these algorithms could achieve $O(n)$ work by
using \emph{compaction} on the set of live nodes at each round. This could reduce their memory footprint, and would be interesting to explore in future work.

\begin{table}[H]
  \centering
  \small
  \renewcommand*{\arraystretch}{1.1}
  %\resizebox{\columnwidth}{!}{%
  \begin{tabular}{l | l l l l}
    \hline
    \textbf{Benchmark} & \textbf{Problem size} & \textbf{Input memory} & \textbf{Tree size} & \textbf{Memory} \\
    \hline
    Spellcheck & $10^6$ strings & $80$MB & 6M & 312MB \\
    Raytracer & $8$M pixels & $192$MB & 24M & 1.3GB \\
    String Hash & $10^8$ chars & $100$MB & 9.4M & 462.5MB \\
    Sequence & $10^6$ elems & $20$MB & 33M & 1.96GB \\
    Tree & $10^6$ nodes & $20$MB & 18.8M & 1.3GB \\
    Filter & $10^7$ elems & $200$MB & 2.48M & 193MB \\
    \hline
  \end{tabular}%
  %}
  \caption{RSP tree sizes and the amount of memory consumed by the RSP tree for each benchmark problem.}\label{tab:memory}
\end{table}

\myparagraph{The cost of garbage collection}  When a self-adjusting computation is discarded, the resulting RSP tree must be destroyed, which also entails removing its read nodes from the reader sets of any modifiables that they read. Table~\ref{tab:gc} shows the runtime of garbage collection for each of the RSP trees for our six benchmark problems compared to the performance of the initial run. Note that for all of the problems other than Raytracer, garbage collection time is at least a factor of $500$ less than the actual computation. For Raytracer, garbage collection is slightly more costly since it has many readers per modifiable and hence has to pay the cost of deletion from the reader sets. Even then, garbage collection takes less than $1\%$ of the time of the initial run.

\begin{table}[H]
\centering
\small
\renewcommand*{\arraystretch}{1.1}
\begin{tabular}{l | c | c }
  \hline
  \textbf{Benchmark} & \textbf{Initial Run} & \textbf{Garbage Collection (32ht)}  \\
    \hline
    Spellcheck & 750ms & 158us  \\
    Raytracer & 323ms & 1.99ms  \\
    String Hash & 61ms & 101us  \\
    Sequence & 464ms & 437us    \\
    Tree & 242ms & 493us        \\
    Filter & 31ms & 31us        \\
    \hline
\end{tabular}
\caption{The cost of garbage collection for each benchmark problem. The initial run is the performance on 32 threads or 32 hyperthreads, whichever is better.}\label{tab:gc}
\end{table}

\subsection{Additional experiments}

Finally, we perform two additional small experiments that measure the effect that data granularity and the sizes of reader sets have on the overall performance of self-adjusting computation. 

\myparagraph{Granularity tradeoffs}
An important consideration when implementing parallel algorithms is careful control of granularity. This is perhaps even more true when implementing self-adjusting algorithms, since the granularity of the data and the functions executed by readers will directly influence the size of the RSP tree and the overhead of modifiables. A larger granularity will lead to lower work and memory overheads. The tradeoff, however, is that if the granularity is too large, updates will slow down, since more irrelevant information will be recomputed when a small piece of the input is updated. Here, we will explore the performance implications and tradeoffs that come from tuning the granularity of our string hashing benchmark. Results are shown in Table~\ref{tab:granularity}.

\begin{table}[H]
  \centering
  \small
  \renewcommand*{\arraystretch}{1.1}
  %\resizebox{\columnwidth}{!}{%
  \begin{tabular}{l l | c c | c c}
  \hline
  \multirow{2}{*}{\textbf{Granularity}} & \multirow{2}{*}{\textbf{Memory}} & \textbf{Run} & \textbf{Update} & \textbf{Run} & \textbf{Update} \\
  & & $p = 1$ & $k = 1$ & $p = 32\text{ht}$ & $k = 10^4$ \\
  \hline
  16 & 1.85GB & 6.8s & 17us & 158ms & 4ms\\
  32 & 925MB & 4.11s & 15us & 95ms & 3.64ms \\
  64 & 462.5MB & 3.03s & 14us & 62ms & 3.12ms \\
  128 & 231.25MB & 2.5s & \textbf{14us} & 44ms & 3.04ms \\
  256 & 115.63MB & 2.3s & 15us & 34ms & 2.88ms \\
  512 & 57.81MB & 2.1s & 18us & 31ms & \textbf{2.83ms}  \\
  1024 & 28.79MB & 2.0s & 24us & 29ms & 3.68ms \\
  2048 & 14.45MB & 2.0s & 39us & 28ms & 5.79ms \\
  \hline
  \end{tabular}%
  %}
  \caption{Memory usage, initial run speed and update speed for various granularities in the hashing benchmark with size $n = 10^8$. Memory denotes memory used by the RSP tree.}\label{tab:granularity}
\end{table}

\noindent As expected, the memory usage and work overhead decreases monotonically as the granularity is increased. The more interesting aspect to look at is the update performance. Note that it is not necessarily the case that the smallest granularity will lead to the fastest updates. Although a smaller granularity means less redundant data is read and recomputed, it also leads to larger RSP trees, which might negate the benefit. The optimal granularity for update speed will therefore be one that balances the tradeoff between reading data and reducing the RSP tree size. For our string hashing benchmark, we observe that the optimal tradeoff occurs at a granularity of 128 characters for single character ($k = 1$) updates, and at 512 characters for larger ($k = 10^4$) updates. This phenomena is explainable by cache line reads.
%Using a granularity of 512 will reduce the depth of recursion, and hence the number of cache misses by about $9$, while reading a chunk of $512$ characters corresponds to $8$ cache lines, which balance out.

\myparagraph{Impact of reader-set size}
Most self-adjusting computations, including all but one of our benchmarks, only have a constant number of readers (often just one) per modifiable. The raytracer benchmark illustrates the effect of having a large number of readers per modifiable, exhibiting a lesser speedup compared to most of the others. Here, we present a small microbenchmark that examines the performance impact of varying the number of readers of a modifiable. In Table~\ref{tab:reader-set-size}, we depict the results of experiment in which $10^6$ workers in parallel each read from a random modifiable and write its value to a unique output destination. We vary the number of modifiables to observe the effect on performance.

\begin{table}[H]
\centering
\small
\renewcommand*{\arraystretch}{1.1}
\begin{tabular}{c | c | c | c}
\hline
\textbf{\# Mods} & \textbf{Readers/Mod} & \textbf{Run} & \textbf{Update}  \\
\hline
$1$     & $10^6$ & 55.1ms & 191ms    \\
$10$    & $10^5$ & 48.9ms & 183ms    \\
$10^2$  & $10^4$ & 47.2ms & 163ms    \\
$10^3$  & $10^3$ & 46.5ms & 130ms    \\
$10^4$  & $10^2$ & 45.1ms & 57.5ms   \\
$10^5$  & $10$   & 38.4ms & 57.0ms   \\
$10^6$  & $1$    & 27.8ms & 44.3ms   \\
\hline
\end{tabular}
\caption{Runtime of the reader-set size microbenchmark for varying numbers of input modifiables. Run denotes the runtime of the initial run, and Update denotes the runtime of a making a dynamic update to every modifiable.}\label{tab:reader-set-size}
\end{table}

\noindent We observe that for the initial run (the \textbf{Run} column), the performance is only marginally impacted as the number of readers per mod varies from $10$ to $10^6$. The exception is when there is only one expected reader per mod, in which case the performance is up to twice as fast as the $10^6$ reader case. This is because of the optimization we perform in  which modifiables with a single reader store that reader inline instead of allocating a linked data structure. We measure the effect on updates (the \textbf{Update} column) by changing the value of all of the modifiables and propagating the result. We observe that when varying from $10$ to $10^6$ reads per modifiable, performance is at most a factor of four slower, or a factor of five slower compared to the one reader case.

  \section{Conclusion}

In this work, we designed, analyzed, and implemented a system
for parallel self-adjusting computation. We showed that a small
set of primitives is sufficient to express self-adjusting
programs that can exploit arbitrary nested parallelism. Compared to previous work, this is the first such system with theoretical bounds on the runtime of the updates.
Our experiments show that the system is capable of producing
dynamic algorithms that both vastly outperform their static
counterparts when performing small to moderately sized
updates, and scale well on multiprocessor machines.

\section*{Acknowledgments}

We thank the anonymous referees for their comments and suggestions. We also thank Laxman Dhulipala for many useful discussions and for providing valuable guidance. This research was supported by NSF grants CCF-1901381, CCF-1910030, and CCF-1919223.

  \clearpage
  
  \bibliographystyle{abbrv}
  \bibliography{ref,umut}
  
  %\clearpage
  
  %\appendix
  
  %{\noindent\bfseries\LARGE APPENDIX}
  
  %\input{proofs.tex}
  
  %\input{bst.tex}

  %\input{framework-appendix.tex}

  %\input{benchmark-descriptions.tex}

  %\clearpage
  
  %\input{plots.tex}
  
\end{document}